\documentclass{article}
\usepackage{ijcai25}

\usepackage{times}
\usepackage{soul}
\usepackage{url}
\usepackage[hidelinks]{hyperref}
\usepackage[utf8]{inputenc}
\usepackage[small]{caption}
\usepackage{graphicx}
\usepackage{amsmath}
\usepackage{amsthm}
\usepackage{booktabs}
\usepackage{algorithm}
\usepackage{algorithmic}
\usepackage[switch]{lineno}

\title{A Systematic Survey on Federated Sequential Recommendation}

\author{
    Yichen Li$^1$, 
    Qiyu Qin$^1$\thanks{Qiyu Qin and Gaoyang Zhu contributed equally to this work.}, 
    Gaoyang Zhu$^{1*}$, 
    Wenchao Xu$^2$, \\ 
    Haozhao Wang$^{1}$,
    Yuhua Li$^1$,
    Rui Zhang$^1$\thanks{Rui Zhang and Ruixuan Li are corresponding authors.},
    Ruixuan Li$^{1\dagger}$
    \affiliations
    $^1$School of Computer Science and Technology, Huazhong University of Science and Technology\\
    $^2$Department of Computing, The Hong Kong Polytechnic University
    \emails
    \{ycli0204, hz\_wang, idcliyuhua, rxli\}@hust.edu.cn, wenchao.xu@polyu.edu.hk, rayteam@yeah.net
}

\begin{document}

\maketitle
\begin{abstract}
     Sequential recommendation is an advanced recommendation technique that utilizes the sequence of user behaviors to generate personalized suggestions by modeling the temporal dependencies and patterns in user preferences. However, it requires a server to centrally collect users' data, which poses a threat to the data privacy of different users. In recent years, federated learning has emerged as a distributed architecture that allows participants to train a global model while keeping their private data locally. This survey pioneers Federated Sequential Recommendation (FedSR), where each user joins as a participant in federated training to achieve a recommendation service that balances data privacy and model performance. We begin with an introduction to the background and unique challenges of FedSR. Then, we review existing solutions from two levels, each of which includes two specific techniques. Additionally, we discuss the critical challenges and future research directions in FedSR.
\end{abstract}

\begin{figure}[t]
    \centering
    \includegraphics[width=\linewidth]{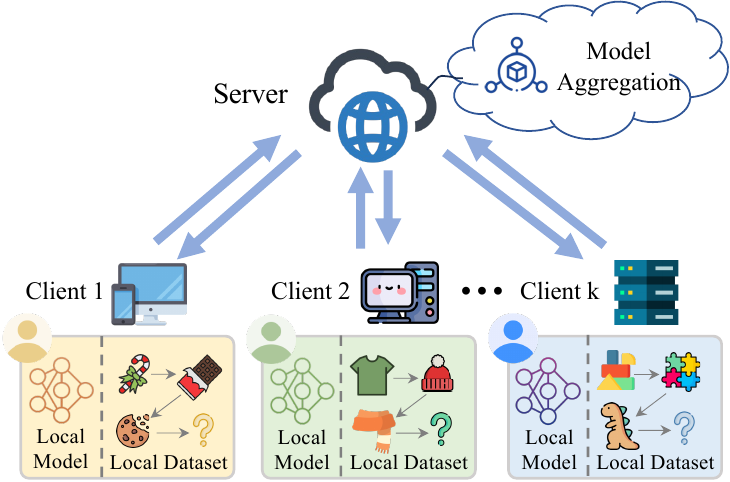}
    \caption{The framework of FedSR. In FedSR, each user participates in the federated training with the local model and dataset. Firstly, the user trains the local model with the private dataset locally and then uploads the model parameters to the server. The server aggregates the local models and broadcasts the global model to the users participating in the next communication round.}
    \label{framework}
    \vspace{-10pt}
\end{figure}
\section{Introduction}
Sequential recommendation has emerged as a pivotal area of research in the field of recommender systems \cite{sr1,sr2}. Unlike traditional recommendation approaches that treat user interactions as independent events, sequential recommendation acknowledges the inherent order and temporal dependencies in user behavior \cite{sr3}. By modeling the sequence of actions or items that users engage with over time, these systems aim to capture the dynamic nature of user preferences and generate more accurate and personalized recommendations. However, in such cases, data collection in central entities, referred to as centralized training, is often infeasible or impractical due to data privacy concerns or country regulations \cite{gdpr}. As a potential remedy to this dilemma, federated learning emerges as a promising approach, enforcing data localization and enabling the distributed training of a globally shared model \cite{li2020federated,mcmahan2017communication}. This framework has achieved remarkable success and has been applied to various fields, such as recommendation systems \cite{intro1} and smart healthcare \cite{flmedical1}.

In recent years, Federated Sequential Recommendation (FedSR) has been proposed to allow users to collaboratively train a shared model locally without breaching their privacy \cite{intro2}. Moreover, it effectively reduces the demand for computing resources on the server while efficiently utilizing the local resources of distributed devices \cite{intro1}. Given the aforementioned advantages, FedSR has gained success in many applications. For instance, \cite{intro3} proposes using pseudo-labeling and secret-sharing techniques to protect privacy while ensuring the performance of federated recommendation systems. \cite{intro4} utilizes Word2Vec to efficiently learn embeddings for users and items, combined with privacy protection mechanisms to achieve cross-device federated recommendation. FedNRM is proposed in \cite{intro5} to design a personalized news recommendation model that combines federated learning and privacy-preserving technologies to provide high-quality recommendation services while protecting privacy. The framework of FedSR is illustrated in Fig.\ref{framework}.

Although many research works investigate integrating FL algorithms with sequential recommendation, several challenges must be solved before FedSR can be scalable. Firstly, data heterogeneity across clients presents a significant challenge. Each user participates in federated training as a participant, and there are significant differences in preferences among individual users. Additionally, the wide variety of items results in sparse patterns for each user, necessitating robust aggregation strategies to ensure that the global model generalizes well. Secondly, communication efficiency is crucial. Compared to ResNet \cite{he2016deep} backbones commonly used in other tasks like image classification, sequential recommendation models are often large in size, resulting in significant communication overhead. Meanwhile, in FL systems, there are often numerous clients with limited resources, requiring reduced communication without sacrificing performance. Thirdly, user privacy requires extra care. While FL keeps data localized, uploading the model updates may still leak sensitive information with malicious attacks, which may include the user's personal information.

While surveys on federated learning and sequential recommendation exist, the existing studies treat the two topics separately. The remaining surveys primarily focus on federated recommendations \cite{fr_survey1,fr_survey2}. As an essential branch of recommendation systems with numerous research works, sequential recommendation merits separate analysis due to its value and significance. Moreover, most existing surveys still concentrate on the classification of federated learning based on heterogeneity and security, neglecting the technical classification of recommendation techniques. This paper proceeds from two levels and distills two mainstream techniques for each (parameter decomposition and LLM foundation for model-level, communication optimization, and aggregation balance for device-level).

Specific contributions of our survey are as follows: 1) We present the first comprehensive survey of recent advancements in federated sequential recommendation, backgrounds, related works and new insights. 2) Our systematic categorization of these methods into two subcategories based on their defining characteristics offers a thorough and structured overview. Based on this, we further elaborate on the existing methods by identifying and discussing two key techniques for each category. 3) We highlight the current challenges and potential future directions in federated sequential recommendations. We intend to shed light on under-researched aspects to spur possible paths within this field.

\section{Preliminaries}
The main objective of this section is to establish a solid foundation for the reader by providing the background and fundamentals of FedSR. We will first analyze the backgrounds of federated learning and sequential recommendation. Then, we discuss the new insights associated with FedRS.

\subsection{Federated Learning}
In recent years, due to growing concerns about privacy leakage, Federated Learning (FL) has been introduced for machine learning across distributed local clients. FL enables multiple users to collaboratively contribute to a global model by exchanging and aggregating model parameters \cite{li2020federated,mcmahan2017communication}. This approach reduces communication overhead and effectively preserves data privacy, as only parameter transfers are involved. Google introduced the foundational FL algorithm, FedAvg, in 2017 \cite{mcmahan2017communication}. Since then, numerous studies have aimed to optimize it further. For instance, FedProx \cite{li2020federated} addresses the heterogeneity issue by adding a regularization term to penalize local model divergence from the global model. MOON \cite{li2021model} enhances local training of different clients by applying contrastive learning at the model level, comparing model representations.

Next, we will outline the federated learning process in three main steps: 1) The server initializes a global model $w^0$ and transmits it to all clients after determining the training tasks and specifying hyper-parameters; 2) In each communication round $t$, a random subset of clients is selected. These clients receive the current global model $w^t$ and perform SGD on their local data. After local training, they upload the updated parameters to the server; 3) During the aggregation phase, the server combines the local models and distributes the updated global model $w^{t+1}$ to the clients participating in the next round. The aggregation is performed as follows:
\begin{equation}
w^{t+1} = \sum_{k \in S_t} \frac{|D_k|}{|D|} w^{t+1}_k.
\end{equation}
where $|D|$ is the total amount of data, $|D_k|$ is the amount of data on client $k$, and $S_t$ is the set of clients participating in the $t$-th round.

\subsection{Sequential Recommendation}
In recent years, with the extensive application of user behavioral data, Sequential Recommendation (SR) has gradually become a key area of research in recommender systems. Its core objective is to predict the next item of interest based on the chronological order of a user’s historical interaction behaviors.
Early studies on sequential recommendation commonly relied on Markov chains to capture the sequential dependencies of user behaviors. For example, FPMC \cite{psr4} combines a Markov chain with a matrix factorization model to simultaneously capture users’ long-term interests and short-term sequential transitions. 
Later, deep learning is extensively used to model user behavior sequences. Convolutional neural networks can be applied to sequential recommendation \cite{psr6} by extracting sequential patterns from users’ historical interactions through convolution operations. 

We divide the main process of sequential recommendation into the following three stages: 1) Input Sequence Construction: Extract behavioral sequences from user interaction logs and represent them as item ID lists or embedding matrices in chronological order; 2) Sequential Pattern Modeling: Based on the backbone model, capture dependencies in user sequences and learn latent feature representations of these sequential patterns; 3) Next-Step Behavior Prediction: Utilize the modeled user sequence features to predict the user’s next item of interest according to the current context or historical records, typically outputting a Top-N recommendation list. By continuously optimizing sequential modeling methods, this process effectively enhances the performance of recommender systems in dynamic environments, thereby providing significant value to personalized services and user experience.

\subsection{New Insights in Sequential Recommendation by Federated Learning}
Compared to training the sequential recommendation model by collecting data centrally, utilizing the FL paradigm brings three major characteristics. Firstly, FL is inherently designed for privacy preservation, as it trains a global model by allowing raw data to remain on the participants' local devices. Compared to other privacy-preserving technologies, FL does not require complex statistical computations for data encryption or the use of perturbations that sacrifice model accuracy for security. FL has been theoretically proven to achieve the performance upper bound of centralized training. Then, Although the performance of FL can be affected by non-IID data, the data heterogeneity in FedSR differs from traditional FL in that it primarily stems from variations in user information among different participants. This variation persists even in traditional sequential recommendations, where the model still needs to integrate and analyze information from different users' raw data. In FedSR, the participants are users, and during the training process, the server also needs to integrate and analyze information from updates uploaded by different users. Last but not least, introducing FL can enhance the system's update efficiency, enabling each user to analyze and upgrade their local model using local resources dynamically. It also allows each user to customize their information security, thereby providing a better recommendation service for users.

\begin{figure}[t]
    \centering
    \includegraphics[width=\linewidth]{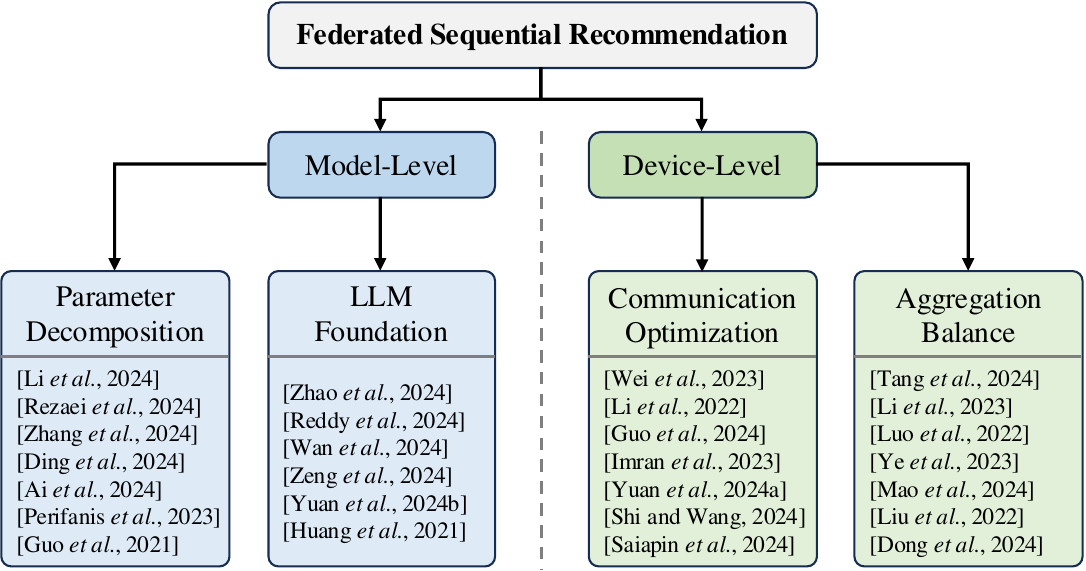}
    \caption{Taxonomy of FedSR. We classify them into two subcategories and four primary techniques, i.e., model-level (parameter decomposition and LLM foundation) and device-level (communication optimization and aggregation balance). Different colors indicate categories, and we list representative works in the boxes.}
    \label{outline}
        \vspace{-10pt}
\end{figure}
\section{Solutions for Federated Sequential Recommendation}
This section outlines a series of targeted solutions designed for FedSR. Considering the research methods of both FL and SR, we categorize the existing studies into model-level and device-level based on the level targeted by the technique. Then, for each level, we separately investigate two major techniques. We illustrate the outline of this survey in Fig.\ref{outline}.

\subsection{Model-Level}
Model-level approaches aim to optimize model architectures and training strategies to balance the trade-offs between privacy preservation, personalized recommendations, and recommendation effectiveness. Specifically, parameter decomposition methods divide model parameters into global and local components, enabling collaborative learning across diverse data sources. This approach allows for the sharing of common information across users while maintaining adaptability to individual behaviors. On the other hand, LLM foundation methods leverage the powerful sequential modeling capabilities of LLMs, combined with the federated learning framework, to enhance the understanding of complex user behavior patterns. These methods improve recommendation accuracy and diversity while preserving user privacy.

\subsubsection{Parameter Decomposition} 
Parameter decomposition has become an effective strategy for addressing the trade-off between model performance and privacy preservation in federated sequential recommendation tasks. Traditional federated learning methods often face challenges when dealing with Non-IID data, heterogeneous models, and computationally intensive tasks. Introducing decomposition in model representation enables efficient collaboration across diverse data sources, tasks, and model components. Through the adoption of global parameters to capture universal patterns across users while leveraging local parameters to adapt to individual user behaviors, this approach ensures privacy protection while optimizing recommendation performance. In sequential recommendation tasks, it guarantees model personalization and enhances robustness.

\begin{figure}[t]
    \centering
    \includegraphics[width=\linewidth]{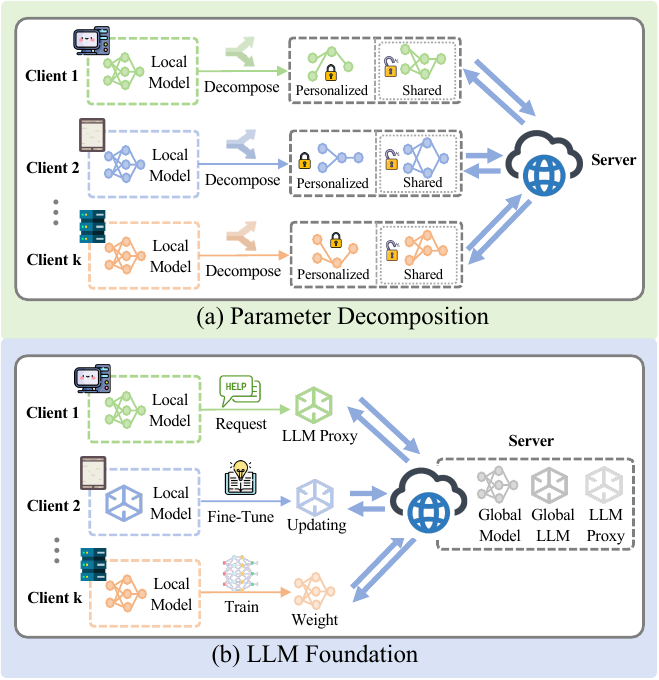}
    \caption{Model-level methods, including parameter decomposition and LLM foundation. Parameter decomposition captures universal patterns across clients while leveraging local parameters to adapt to local dataset, and LLM foundation guides knowledge retention and parameter updates by integrating knowledge from LLM.}
    \label{device-level}
        \vspace{-10pt}
\end{figure}
Several studies aim to balance generalization and personalization. For example, \cite{pd1} proposes a semi-global modeling framework that dynamically optimizes personalization and generalization through alternating local training and global aggregation. On the client side, the model captures personalized features using local data; on the server side, aggregated model information updates the semi-global model, preserving global knowledge. This EM-like optimization strategy groups user behaviors to generate sub-models that align with different user group preferences. Experimental results demonstrate that during the personalized training phase, the proposed method achieves up to a 21.9\% performance improvement on the ML-latest dataset using its two-stage training framework. This improvement is particularly significant in datasets with highly heterogeneous user preferences. The decomposition strategy between semi-global and personalized components significantly enhances the model's performance in sequential recommendation tasks. Another approach to addressing the generalization-personalization trade-off leverages knowledge distillation techniques. FedDist-POIRec \cite{pd2} extracts knowledge from locally trained models in the form of soft labels, which are softened representations of model parameters. These soft labels are then uploaded to the server for aggregation, avoiding the direct parameter synchronization common in traditional federated learning. This method further improves the model's adaptability to heterogeneous data distributions. 

Parameter decomposition has also demonstrated strong applicability in multi-scenario and cross-domain recommendation tasks. FedDCSR \cite{pd3} proposes a cross-domain recommendation framework based on disentangled representation learning to address the challenges of privacy preservation and cross-domain knowledge sharing in cross-domain recommendation tasks. Disentangling domain-shared and domain-specific representations effectively captures user behavior characteristics across different domains. On the client side, local models focus on capturing domain-specific interests, while on the server side, the aggregation of domain-shared representations enhances the global model's generalization capability. Experimental results show that FedDCSR significantly outperforms all baselines across various metrics, highlighting the critical role of disentangled representation learning and contrastive information strategies in capturing both intra-domain and inter-domain user preferences. This provides an effective solution to the problem of cross-domain knowledge transfer. For multi-task scenarios, \cite{pd4} proposes a personalized federated learning framework. Through a multi-task aggregation strategy, the framework balances multiple scenarios within the global model. Its core lies in introducing personalized modules for each task, where local models adapt to user needs in specific scenarios while sharing global knowledge for inter-task collaboration.

In addition to multi-scenario and cross-domain research, personalization optimization and dynamic adaptation have emerged as critical directions for parameter decomposition methods. FMLRec \cite{pd5} introduces a meta-learning framework that captures shared characteristics among users through global meta-training while utilizing fast fine-tuning to adapt to local user behavior data. This method performs exceptionally well in cold-start scenarios. The dynamic adaptation capability of meta-learning demonstrates greater flexibility and robustness in highly heterogeneous data scenarios.

Parameter decomposition methods have also shown unique value in specialized tasks. FedPOIRec \cite{pd6} incorporates social influence modeling to capture social relationships and user interaction behaviors, further enhancing the model's ability to personalize point-of-interest recommendations. Regarding sequential modeling capabilities, PREFER \cite{pd7} focuses on point-of-interest recommendations in edge scenarios. Lightweight model designs and efficient parameter update strategies significantly optimize recommendation performance. Experimental results indicate that these methods not only improve recommendation metrics but also exhibit strong adaptability in heterogeneous data scenarios.

\subsubsection{LLM Foundation}
Beyond parameter decomposition-based approaches, a series of methods leveraging Large Language Models (LLM) have been developed to address the challenges of federated sequential recommendation. These methods combine the powerful sequential modeling capabilities of LLMs with the privacy-preserving mechanisms of federated learning, providing innovative solutions for recommendation systems. By leveraging LLMs' deep semantic modeling of user behavior sequences, item features, and contextual information, they significantly enhance recommendation performance. Meanwhile, the federated learning framework ensures privacy protection by avoiding the direct transmission of sensitive data, achieving a balance between privacy preservation and personalized recommendations. In recent years, researchers have proposed various innovative techniques focusing on deep modeling, retrieval-augmented generation, reinforcement learning, and knowledge enhancement, driving the practical application of federated sequential recommendation in complex tasks.

The introduction of LLM in federated sequential recommendation research has significantly enhanced the ability to capture complex patterns in user behavior and generate more accurate personalized recommendations. PPLR is proposed in \cite{lf1} to combine the powerful semantic representation capabilities of LLMs with the privacy-preserving mechanisms of federated learning, achieving dual optimization of recommendation performance and user privacy protection. This method leverages pre-trained LLMs to generate high-dimensional semantic representations of user behaviors, item descriptions, and contextual information, enabling richer feature expression. Furthermore, it enhances the generalization capability of global models and the personalization of local models through parameter aggregation under the federated learning framework and personalized fine-tuning. Experimental results demonstrate that this method outperforms other privacy-preserving federated recommendation methods across three datasets: Games, MicroLens, and Book while achieving performance close to centralized LLM-based approaches. This establishes a groundbreaking solution for the future development of federated recommendation systems. In addition to LLMs, Transformer and lightweight deep learning architectures have shown significant potential in federated sequential recommendation. 

The study in \cite{lf2} introduces the transformer architecture into federated recommendation systems. By leveraging multi-head attention mechanisms, the model effectively captures complex patterns in user behavior sequences. At the same time, its lightweight Transformer design makes it well-suited for federated learning scenarios with limited communication resources, enhancing global user behavior modeling capabilities while maintaining privacy protection. Another study, Fed-AttGRU \cite{lf3}, combines attention mechanisms with GRU, focusing on the most important parts of user behavior sequences and modeling both short-term preferences and long-term interests. Experimental results indicate that Fed-AttGRU performs exceptionally well in long-sequence modeling and highly heterogeneous data scenarios, highlighting its value in federated recommendation research.

For cold-start and sparse data scenarios, strategies combining retrieval and generation have provided practical solutions for federated recommendations. GPT-FedRec \cite{lf4} proposes a retrieval-augmented generation framework, which generates more contextually relevant recommendations through a hybrid retrieval and generation strategy. The federated learning framework protects user data during local training while integrating retrieval and generation modules, which improves system performance in sparse data and cold-start scenarios. GPT-FedRec demonstrated remarkable improvements across six datasets and four evaluation metrics, achieving average increases of 36.12\% in Hit@5, 29.88\% in NDCG@5, 45.44\% in Hit@10, and 36.56\% in NDCG@10. his study highlights the potential of combining retrieval and generation techniques in the federated recommendation and further advances the application of such systems to address cold-start challenges.

Regarding knowledge enhancement, FELLAS \cite{lf5} proposes a framework that leverages LLMs as external services to enhance federated sequential recommendations. By interacting with LLMs, the framework provides local models with rich semantic information and contextual features, enabling the recommendation model to capture better user behavior characteristics in sparse data and long-tail scenarios. By optimizing the frequency of LLM calls and communication strategies, FELLAS effectively reduces local resource consumption and avoids the challenges of directly deploying complex LLM models, showcasing the potential of external knowledge enhancement in FedSR.

Furthermore, LLM technologies have been widely applied in dynamically optimized recommendation tasks. \cite{lf6} introduces deep reinforcement learning into federated learning. By locally training reinforcement learning agents, this method models the dynamics and long-term preferences of users' schedules. Meanwhile, global parameter sharing integrates knowledge across multiple users. This combination of reinforcement learning and federated learning demonstrates advantages in dynamic recommendation tasks and further validates the applicability of federated learning frameworks in complex recommendation scenarios.

\subsection{Device-Level}
Device-level approaches aim to address the heterogeneity challenges in the federated sequential recommendation by enhancing the performance and adaptability of recommendation systems through the optimization of model aggregation strategies. To better capture user behavior characteristics, optimizing device selection and device clustering mechanisms becomes crucial. These methods effectively improve both model training efficiency and recommendation performance by focusing on optimizations at the device level. In addition to optimizing device selection and clustering mechanisms to capture user behavior features better, reducing communication overhead is also a vital component. Optimizing communication strategies can effectively alleviate the communication burden between clients and servers, thereby accelerating the global model update speed and reducing latency and resource consumption during the training process.

\subsubsection{Communication Optimization}
Communication optimization primarily focuses on reducing the communication overhead between clients and servers, enhancing system operational efficiency, and decreasing overall communication costs. In addition to reducing communication overhead through techniques such as sparsification, which involves transmitting only critical parameter updates and compressing model updates, some communication optimization strategies also balance the reduction of overhead with the protection of user privacy, which is especially important in distributed environments. Firstly, in the KG-FedTrans4Rec framework \cite{co2}, the authors optimize communication efficiency through an edge device aggregation mechanism. In this model, edge devices not only perform local training but also undertake information aggregation tasks, thereby reducing the frequency of model updates and the volume of data transmission.

\begin{figure}[t]
    \centering
    \includegraphics[width=\linewidth]{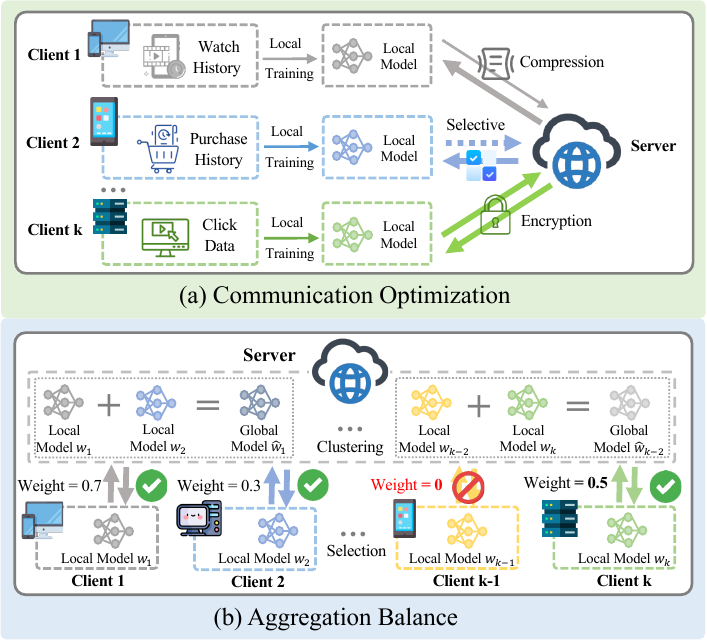}
    \caption{Device-level methods, including communication optimization and aggregation balance. Communication optimization focuses on reducing the communication cost between clients and the server, and the aggregation balance often strategically aggregates models to maximize the optimization of the global model.}
    \label{device-level}
        \vspace{-10pt}
\end{figure}
In \cite{co1}, the authors propose a method to effectively reduce the amount of communication data by utilizing low-rank tensor projections and device clustering strategies. This approach models users' long-term preferences by projecting user behavior data into a low-rank space and transmitting only smaller updates during each communication round. Similarly, researchers in \cite{co3} address communication challenges in horizontal federated recommendation systems by applying model compression techniques. By quantizing and sparsifying model parameters, they significantly decrease the volume of transmitted gradient data, thereby improving communication efficiency. Furthermore, the proposed strategy, based on gradient optimization for model compression, further reduces the size of transmitted data by sharing compressed model updates, optimizing overall communication efficiency.

Beyond data compression, selective data transmission can also be employed. ReFRS \cite{co5} significantly reduces communication data volume by transmitting only the encoder parameters of the client-embedded model to the server. The server side employs asynchronous dynamic clustering methods and semantic samplers to extract low-dimensional embedding representations and group similar client models, thereby avoiding sensitivity to Non-IID data and reducing communication overhead. ReFRS demonstrates superior memory and computational efficiency, with embeddings using only 270KB compared to FedAvg and FedFast, which require 2.43MB-3.78MB, and batch computation time reduced from 45 minutes for FedAvg and FedFast to 3 seconds. Furthermore, in PTF-FSR \cite{co6}, the authors take it a step further by proposing a parameter-free federated learning framework that replaces model parameters with user-generated sequential data for knowledge transfer, completely eliminating the high communication overhead associated with parameter transmission. This approach is suitable for the efficient operation of large-scale complex models, with PTF-FSR’s communication cost ranging from only 1.2KB to 2.4KB, significantly lower than traditional federated methods such as Fed-SASRec, which range from 1.6MB to 14.8MB, reducing communication overhead by more than an order of magnitude.

Additionally, some communication optimization methods offer the dual benefits of reducing communication overhead and enhancing privacy protection. For example, the FedSeqRec framework \cite{co4} employs local differential privacy techniques to encrypt and compress model parameters, thereby optimizing both user data privacy and communication efficiency. After local model training and parameter updates, clients encrypt the parameters using LDP techniques before sending them to the central server. This ensures the security and compressibility of the transmitted data, alleviating the communication burden. Similarly, the SeqMF model \cite{co7} incorporates the QHarmony mechanism, which significantly reduces communication overhead by transmitting sparse gradients with quantization perturbations. This mechanism selectively transmits partially quantized gradients and associated metadata, minimizing the amount of data transmitted while preserving training effectiveness and privacy.

\subsubsection{Aggregation Balance}
Aggregation balance addresses the challenges of uneven data distribution and heterogeneous device characteristics in federated learning by dynamically adjusting each device's contribution to the global model. This ensures that devices with more critical or representative data play a more significant role in the aggregation process. Specific strategies include assigning weights to local models based on data importance and clustering devices with similar data distributions. These methods effectively mitigate data sparsity issues and enhance the model's ability to provide personalized recommendations.

Some methods primarily optimize recommendation performance through client selection. For example, FedGST \cite{xz1} optimizes recommendation performance through a device-level client selection mechanism. This method uses influence functions to assess each client's contribution to model training dynamically. In each training round, the server selects high-contribution clients based on their contribution values to participate in the next round of training, thereby improving model training efficiency and recommendation performance. DistVAE \cite{xz7} optimizes recommendation performance through a contribution-based client selection method and clustering strategy. This framework evaluates each client's computational capability and training contribution, selects high-contribution clients to participate in training, and clusters clients based on device resources and behavioral characteristics to reduce communication overhead and enhance recommendation performance, making it suitable for large-scale distributed recommendation systems. CF-FedSR \cite{xz8} combines adaptive client selection, client clustering-based sampling, fairness-aware model aggregation, and personalized recommendation modules to optimize recommendation performance. This framework optimizes participating clients through an adaptive selector, groups and proportionally samples similar clients based on clustering methods, employs weighted strategies for model aggregation, and integrates global models with locally fine-tuned models to achieve personalized recommendations, thereby significantly enhancing communication efficiency, fairness, and recommendation performance while also reducing communication consumption. For instance, on the Beauty dataset, CF-FedSR improves HR@10 by 8.90\% and NDCG@10 by 15.46\% compared to FedAvg while reducing communication rounds by approximately 10.67\% compared to FedAvg.

Other methods primarily focus on enhancing recommendation performance through client clustering. For instance, CPF-POI \cite{xz2} employs an adaptive clustering mechanism to optimize recommendations. This framework utilizes a gate control strategy based on Gumbel-Softmax sampling to dynamically adjust client clustering schemes, grouping similar clients together to facilitate knowledge sharing and prevent negative transfer. This balance enhances both knowledge sharing and personalized recommendations. In CPF-GCN \cite{xz3}, a cluster-driven approach is used to optimize recommendation performance. The server clusters users based on their embeddings and selects representative clients from each cluster proportionally for model updates. This strategy not only reduces communication costs but also improves the model's generalization and personalized recommendation capabilities.

Additionally, FCLUB \cite{xz4} proposes a method that combines stage-wise clustering detection algorithms with asynchronous communication protocols to boost recommendation performance. Clients generate local clustering information based on users' historical interaction data, dynamically adjust local connection graphs to form multiple clusters, and upload this information to the global server for global clustering, maximizing collaborative effects. Furthermore, SFL \cite{xz5} optimizes recommendation performance by incorporating semantic information and client clustering strategies. This framework clusters clients based on user behavior's semantic features, ensuring similar users are grouped to promote knowledge sharing and reduce interference from unrelated data. It also reduces communication overhead by transmitting perturbed semantic information. On the NYC dataset, using the SASRec model, SFL improves NDCG@5 from 0.2789 to 0.4492 compared to FedAvg, a 60.8\% increase. In single-round computations, server-side computation time decreases from 22.86 seconds with FedAvg to 9.28 seconds with SFL.

Moreover, spatial and temporal information can be leveraged for clustering. For example, SCFL \cite{xz6} adopts a space-time consistency-based clustering strategy to optimize recommendation performance. This method enhances user collaboration through hierarchical aggregation strategies and edge device clustering, utilizes trajectory optimization modules to extract deep behavioral patterns, and facilitates information sharing and aggregation through edge devices. This approach improves personalized recommendation capabilities while reducing computational overhead. On the NYC dataset, using the SASRec model, SCFL achieves an NDCG@10 of 35.42, approximately 40\% higher than FedProx’s 25.26. It also significantly reduces server-side computation time, decreasing from 16.35 minutes with FedProx to 2.07 minutes with SCFL.

\section{Future Directions}
Although there is a lot of existing research, there are still challenging new research directions in the deployment of FedSR to be discussed as follows.
\begin{itemize}
    \item \textit{Personalized Strategy}: Because different users have varying preferences, it is inadequate to provide the same recommendation service to all users in FedSR. Federated sequential recommendation can investigate personalized strategies to optimize recommendation performance by analyzing user preferences and their individual understanding of items.
    \item \textit{LLM Foundation}: Recently, LLM has gained significant attention due to its outstanding capabilities, leading researchers to explore its incorporation into the sequential recommendation. Nevertheless, it presents a considerable challenge in FedSR. In these environments, distributed devices must communicate frequently to exchange knowledge from ongoing tasks, while the server needs to efficiently extract new insights from the LLM. Problems concerning communication efficiency and training resources can hinder model convergence. Future research should investigate novel methods for LLM-based FedSR to address these challenges.
    \item \textit{Multi-Modal Fusion}: In the research of recommendation systems, enhancing knowledge representation is commonly employed to improve model performance. In existing research on FedSR, the issue of multi-modality has been relatively under-considered. However, in addition to user-item interaction records, user reviews, user personal information, and item information can all serve as auxiliary information to enhance model performance. Future research could focus more on how to integrate information from different modalities while protecting the privacy therein to improve model performance.
    \item \textit{Cold-Start Adaptation}: In the context of shopping recommendations, some items may have fewer interaction records and present a cold-start problem for FedSR systems. Traditional methods may struggle to provide accurate recommendations when there is limited or no historical interaction data. Future research should explore adaptive strategies to handle cold-start scenarios, such as leveraging transfer learning techniques to utilize knowledge from similar cases or incorporating expert knowledge to initialize recommendations.
    \item \textit{Dynamic Preference}: User preferences are typically dynamic and ever-changing. Federated sequential recommendation needs to capture such dynamic changes to provide more accurate recommendations. Federated continual learning has attracted growing interest by enabling distributed devices to collaboratively learn novel concepts from streaming training data while avoiding knowledge forgetting of previous tasks. Future research could consider combining CL and FedSR to provide dynamic recommendation services.
\end{itemize}

\section{Conclusion}
We provide a comprehensive survey of the federated sequential recommendations. First, we begin with an introduction to the motivation for integrating the FL paradigm into the sequential recommendation and related works. Then, we categorize existing methods into two categories with four primary techniques. Finally, we point out the future directions of federated sequential recommendation. We expect this survey to provide an up-to-date summary of recent work and inspire new insights into the sequential recommendation field.

\section*{Acknowledgments}
This work is supported by the National Key Research and Development Program of China under grant 2024YFC3307900; the National Natural Science Foundation of China under grants 62376103, 62302184, 62436003 and 62206102; Major Science and Technology Project of Hubei Province under grant 2024BAA008; and Hubei Science and Technology Talent Service Project under grant 2024DJC078.

\bibliographystyle{named}
\bibliography{ijcai25}

\end{document}